\documentclass[a4paper,12pt]{article}
\input{epsf}
\newcommand{\bcen}{\begin{center}}
\newcommand{\ecen}{\end{center}}
\newcommand{\btab}{\begin{tabular}}
\newcommand{\etab}{\end{tabular}}
\newcommand{\bdes}{\begin{description}}
\newcommand{\edes}{\end{description}}

\newcommand{\beq}{\begin{equation}}
\newcommand{\eeq}{\end{equation}}
\newcommand{\bea}{\begin{eqnarray}}
\newcommand{\eea}{\end{eqnarray}}
\newcommand{\non}{\nonumber}

\newcommand{\half}{\frac{1}{2}}
\newcommand{\bary}{\begin{array}}
\newcommand{\eary}{\end{array}}
\newcommand{\be} { \mbox{\boldmath $e$}}

\newcommand{\bn} { \mbox{\boldmath $n$}}

\newcommand{\bs} { \mbox{\boldmath $s$}}

\newcommand{\bx} { \mbox{\boldmath $x$}}

\newcommand{\bF} { \mbox{\boldmath $F$}}

\newcommand{\bI} { \mbox{\boldmath $I$}}

\newcommand{\bR} { \mbox{\boldmath $R$}}

\newcommand{\dou}{\partial}

\newcommand{\D}[1]{\mbox{d}{#1}} 

\newcommand{\prn}[1] {(\ref{#1})}

\newcommand{\fig}[1]{fig.~\ref{#1}}
\newcommand{\Fig}[1]{Fig.~\ref{#1}}

\usepackage{epsfig}
\usepackage{harvard}
\usepackage{amsfonts}
\usepackage{eucal}
\newcommand{\mylabel}[1]{\label{#1}}

\newcommand{\mytitle}{ Size-dependent Rigidities of  Nanosized Torsional Elements}
\begin{document}

\date{}
\baselineskip=18pt

\title{\sl \Large \mytitle}
\author{Vijay B.~Shenoy\footnote{ Fax: +91-512-597408, e-mail:
{\tt vbshenoy@iitk.ac.in}}  \\ 
Department of Mechanical Engineering \\
Indian Institute of Technology, Kanpur \\ UP 208016, India}
\maketitle

\begin{abstract}
A theory for the prediction of the size dependence of
torsional rigidities of nanosized structural elements is developed. It
is shown  that, to a very good approximation, the torsional rigidity $(D)$ of a nanosized bar differs from the
prediction of standard continuum mechanics $(D_c)$ as $(D-D_c)/D_c = A
h_0/a$ where $A$ is a non-dimensional constant, $a$ is the size scale of
the cross-section of the bar and $h_0$ is a material length equal to
the ratio of the surface elastic constant to the bulk elastic
constant. The theory developed is compared with direct atomistic
calculations (``numerical experiment'') of the torsional rigidity bars made of several FCC metals
modeled using the embedded atom method. Very good agreement is
obtained between theory and simulation. The framework presented here
can aid the development of design methodologies for nanoscale structural
elements without the need for full scale atomistic simulations.
\end{abstract}

\section{Introduction}

The demand for smaller and faster devices have encouraged
technological advances resulting in the ability to manipulate matter
at micro- and nanoscales that have enabled the fabrication of
micro/nanoscale electromechanical systems. While MEMS technology is now
a well established area, nanoelectromechanical systems
(NEMS) have recently made their appearance in literature (see, for
example, \citeasnoun{Roukes2000}). A key feature of NEMS that make
them attractive are their high fundamental frequencies while affording
small force constants. Nanosized bars and tubes are produced in a
variety of ways with materials such as SiC, MoO$_3$ and C (carbon
nanotubes) (see \citeasnoun{Yakobson1997}, \citeasnoun{Terrones1999},
\citeasnoun{Sheehan1996}). These nanosized elements have found many
technological uses, for example, as  probes of  scanning probe microscopes
\cite{Dai1996}, in altering properties of bulk materials in the form
of whisker additions \cite{Kuzumake1998} etc.

In the recent past, several groups have reported studies on the
mechanical behaviour nanosized bars and
nanotubes. \citeasnoun{Wong1997} performed experiments on SiC beams
while \citeasnoun{Poncharal1999}, and more recently
\citeasnoun{Gao2000} have reported experiments where the elastic
modulus of carbon nanotubes is measured using dynamic techniques. A
careful study of these reports show that the elastic moduli of such
nanosized structural elements depend on their size. Attempts to
explain the size dependent behaviour has been through direct atomistic
computer simulation of these structures
\cite{Robertson1992,Garg1998a,Garg1998b}. These studies have reported
size-dependent elastic moduli computed from atomistic simulations.

\citeasnoun{Miller2000} developed a simple model to explain the
size dependence of the elastic rigidities of nanosized structural
elements. These size dependences were attributed to the heterogeneities
in the atomic environments introduced by the bounding free surfaces of the structural
elements. Thus, as the size of the structure becomes smaller the
presence of surfaces have to be accounted for in modeling
strategies. Developing this premise, \citeasnoun{Miller2000} showed
 that the differences between the rigidities  $(D)$ of these small
elements and those predicted by continuum mechanics ($D_c$) can be
expressed as
\bea
\frac{D-D_c}{D_c} = A \frac{h_0}{a} \label{miller}
\eea
where $A$ is a nondimensional constant that depends on the geometry of
the structure, $a$ is the size scale of the
structural element (for example, the cross-sectional width of a bar),
and $h_0$ is a material length that is the ratio of the surface
elastic constant of the bounding surfaces of the structure and the
bulk elastic constant of the material. Thus, the size dependence of
the rigidities can be predicted by obtaining the material parameter
$h_0$ and the nondimensional constant $A$. Typically, $h_0$ can be
obtained from a small atomistic simulation and $A$ can be calculated
analytically. Thus the need for full scale atomistic simulations of
structures (which is an expensive proposition) is
obviated. \citeasnoun{Miller2000} applied this model to study the
elastic properties of nanosized bars, plates and beams and demonstrated the
strength of the model by comparison with direct atomistic calculations.

An important mode of deformation of bar-like structures is torsion;
for example, the probe of a scanning probe microscope is subjected to
bending and torsion. It is therefore important to develop a model for
the torsional rigidities of nanosized elements; this is the aim of the
present paper. An augmented
continuum theory of torsion accounting for the presence of free
surfaces is developed and the size dependence of the rigidity is
derived analytically. A perturbative scheme is developed for
solving the resulting boundary value problem which provides a
simple method to evaluate the size dependence of the torsional rigidity. It is shown
that, in general, the size dependence of the rigidity $D$ is of the
form 
\bea 
\frac{D-D_c}{D_c} = A \frac{h_0}{a} + B \left(\frac{h_0}{a}\right)^2 +
\ldots  
\eea 
where $A$, $B$, etc.~are
non-dimensional constants and  $h_0 = S/G$ where $S$ is the
surface shear modulus and $G$ is the bulk shear modulus.  The
perturbative method also provides a general
framework for the calculation of the nondimensional constants
$A,B,\ldots$ that depend only on the geometry of the cross section of the
bar.  For the case of square bars of side $2a$, the
theory provides that size dependence of torsional rigidity to be very
well approximated by 
\bea 
\frac{D-D_c}{D_c} \approx 4 \frac{h_0}{a}.
\eea 
These theoretical results are then compared with direct atomistic
simulations (which serve as numerical experiments) of torsion of square bars of FCC metals. The methodology
required  for the atomistic simulation of torsion is also developed
here. The agreement theory and numerical experiment (atomistic
simulations) is excellent.

The paper is organised as follows. The next section contains the
augmented theory of torsion. Section \ref{Simulations} contains the
details of simulation methodologies, the results of which are reported
and discussed in section \ref{Results}. The paper in concluded in
section \ref{Conclusions} in which several important directions of
future work are identified.

\section{Theory}
\label{Theory}
\subsection{Augmented Continuum Theory}
It was shown by \citeasnoun{Miller2000} that the elastic properties of
nanoscale structural elements  such as plates, bars and beams can be
explained using a augmented continuum theory that accounts for the
energetics of deforming inhomogeneities such as surfaces and corners,
the effects of which are significant when the length scale of the
structure approaches the atomic scale. Several authors had previously
utilized continuum theories of solids with surface
effects \cite{Gurtin1975,Rice1981,Cammarata1994} to
study a variety of problems ranging from diffusive cavity growth in
stressed solids to stability of stressed epitaxial films. The
formulation outlined in \citeasnoun{Miller2000} is briefly
recapitulated here for the sake of completeness and to set the notation.

The body $\cal B$, described by coordinates $x_i$, considered in the
augmented continuum theory is bounded by a surface $\cal S$. It is assumed
that the surface $\cal S$ is piecewise flat (this assumption eliminates the
need to consider contravariant and covariant components of surface
tensors) and is described by coordinates $x_\alpha$ for each flat
face. The bulk stress tensor in the body ${\cal B}$ is denoted by
$\sigma_{ij}$ and the surface stress tensor by $\tau_{\alpha
\beta}$. Mechanical equilibrium of a bulk material element implies that the bulk
stress tensor satisfies (with no body forces)
\bea
\sigma_{ij,j}  = 0 \mylabel{bulkeq}.
\eea
Equilibrium  of a surface element necessitates that 
\bea
\tau_{\alpha \beta,\beta} + f_\alpha & = & 0 \mylabel{surfteq} \\
\non \\
\tau_{\alpha \beta} \kappa_{\alpha \beta} & = & \sigma_{ij} n_i n_j
\mylabel{surfneq} 
\eea
where $n_i$ is the outward normal to the surface, $f_\alpha$ is the negative of the tangential component of
the traction $t_i = \sigma_{ij} n_j$ along the $\alpha$ direction of
surface $\cal S$, and $\kappa_{\alpha \beta}$ is the surface curvature
tensor. The assumption of the piecewise flat surfaces implies that the
surface curvature vanishes everywhere along the surface except at
corners and edges which have to be treated separately. It
must be noted that the assumption of piecewise flat surface is merely
for the sake of mathematical simplicity; the present theoretical
framework is valid for curved surfaces as well. 

The kinematics of the body is described by the displacement field $u_i$
defined at every point in the body. The strain tensor
$\epsilon_{ij}$ in the body is obtained using a small strain formulation as
\bea
\epsilon_{ij} = \half\left(u_{i,j} + u_{j,i} \right). \mylabel{bulkstr}
\eea
The surface strain tensor $\epsilon_{\alpha \beta}$ is derived from
the bulk strain tensor $\epsilon_{ij}$ such that every material fibre on
the surface has the same deformation whether it is treated as a part
of the surface or as a part of the bulk, i.~e., the surface strain
tensor is compatible with the bulk strain tensor. 

The final ingredient of the augmented continuum theory is the
constitutive relations that relate the stresses to strains. The bulk
is considered to be an anisotropic linear hyperelastic solid with
a free energy density $W$ defined as
\bea
W(\epsilon_{ij}) = \half C_{ijkl} \epsilon_{ij} \epsilon_{kl} \mylabel{hyperelas}
\eea
and the stresses are derived as
\bea
\sigma_{ij} = \frac{\dou W}{\dou \epsilon_{ij}} = C_{ijkl}\epsilon_{kl} \mylabel{bulkelas}
\eea
where $C_{ijkl}$ is the bulk elastic modulus tensor. In this
framework the bulk free energy vanishes for the unstrained solid.
Constitutive relations for the surface stress tensor are more
involved. The surface stress tensor is related to the surface energy $\gamma$
as
\bea
\tau_{\alpha \beta} = \gamma \delta_{\alpha \beta} + \frac{\dou
\gamma} {\dou \epsilon_{\alpha \beta}} , \mylabel{taugamma}
\eea
a relation which is generally attributed to Gibbs
\cite{Cammarata1994}. The surface stress tensor can be expressed as a
linear function of the strain tensor as
\bea
\tau_{\alpha \beta} = \tau^0_{\alpha \beta} + S_{\alpha \beta \gamma
\delta} \epsilon_{\gamma \delta} \mylabel{surfelas}
\eea
where $\tau^0_{\alpha \beta} $ is the surface stress tensor when the
bulk is unstrained (obtained from \prn{taugamma} with
$\epsilon_{\alpha \beta} = 0$) and $S_{\alpha \beta \gamma \delta}$ is
the {\em surface elastic modulus tensor}.
This is an important quantity in that the size dependence of elastic
properties will be shown to be determined  by the ratio of a surface
elastic constant and the bulk elastic constant. The constitutive constants $C_{ijkl}$ and $S_{\alpha
\beta \gamma \delta}$ are external to the augmented continuum theory;
in this paper these are determined from atomistic models of the
materials considered.

\subsection{Augmented Theory of Torsion of Bars}
The continuum theory of torsion of bars is a part of classical theory
of elasticity attributed to St.~Venant and Prandtl and is treated in
much detail by \citeasnoun{Sokolnikoff1956}. In this section we
develop a theory of torsion based on the augmented theory of last
section which includes surface effects. Corner effects are neglected
in this treatment.

\begin{figure}
\centerline{\input{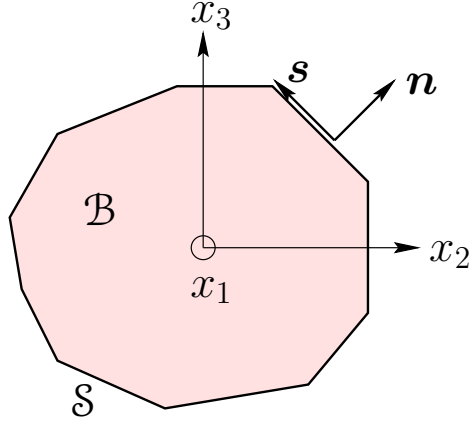}}
\caption{Cross section of bar considered in the augmented continuum theory of torsion.}
\mylabel{section}
\end{figure}

\Fig{section} shows the cross-section of the bar $\cal B$ bounded by a
surface $\cal S$. Attention is restricted to simply connected
cross-sections for the sake of simplicity. The outward normal to $\cal
S$ is denoted by $\bn$ and $\bs$ is the tangent vector to $\cal S$
with  $\bn \cdot \bs = 0$. Application of a torque $T$ to the bar
produces a twist per unit length $\alpha$; the aim of the analysis is
to obtain a relationship between $T$ and $\alpha$. The kinematics is
described by the displacement field as
\bea
u_1(x_1,x_2,x_3) & = & \alpha \phi(x_2,x_3) \non \\
u_2(x_1,x_2,x_3)  & =  & - \alpha x_1  x_3 \mylabel{displ} \\
u_3(x_1,x_2,x_3)  & = &   \alpha x_1 x_2  \non 
\eea
where $\phi$ is the warping function. The only nonvanishing strain
components  derived from these displacements are
\bea
\epsilon_{12}(x_2,x_3)  =  \frac{\alpha}{2} \left(\frac{\dou \phi}{\dou
x_2} - x_3 \right), \;\;\;\;\; \epsilon_{13}(x_2,x_3)  =  \frac{\alpha}{2} \left(\frac{\dou
\phi}{\dou x_3} + x_2 \right) \mylabel{stdisp}
\eea
The bulk material is assumed to be a linear elastic solid and thus the
bulk stresses are related to the strain via
\bea
\sigma_{12} = 2 G \epsilon_{12}, \;\;\;\;\; \sigma_{13} = 2 G
\epsilon_{13} \mylabel{shearcons}
\eea
where $G$ is an appropriate shear modulus (for example, if the `1'
direction corresponds to the $\langle 100 \rangle$ direction in an FCC
crystal, $G = C_{44}$). The only nontrivial equilibrium equation in
the bulk is
\bea
\frac{\dou \sigma_{12}}{\dou x_2} + \frac{\dou \sigma_{13}}{\dou x_3}
= 0. \mylabel{ntequil}
\eea
Eqn.~\prn{ntequil} is identically satisfied on the introduction of a
stress function $\psi$ (after Prandtl) such that
\bea
\sigma_{12} = G \alpha \frac{\dou \psi}{\dou x_3}, \;\;\;\;\;\; \sigma_{13} =
- G \alpha \frac{\dou \psi}{\dou x_2}. \mylabel{stfunc}
\eea
On substituting the expressions for the stresses \prn{stfunc} in
\prn{shearcons} and using \prn{stdisp} it is found that the stress
function satisfies
\bea
\frac{\dou^2 \psi}{\dou x^2_2} + \frac{\dou^2 \psi}{\dou x^2_3} = -2
\eea
in the bulk.

The surface is parametrised by the coordinate $x_1$ and distance $s$
measured along the tangent vector $\bs$. The pertinent component of surface stress is $\tau_{1s}$ where $s$
stands for the $\bs$ direction in \fig{section} and satisfies the 
surface equilibrium equation \prn{surfteq}
\bea
\frac{\dou \tau_{1s}}{\dou s} + f_1 = 0. \mylabel{tauequil}
\eea
The second surface equilibrium condition \prn{surfneq} is identically
satisfied since $\kappa_{\alpha \beta} = 0$. The ``surface body force'' $f_1$ is given as
\bea
f_1 = - t_1 & = & -(\sigma_{12} n_2 + \sigma_{13} n_3) \non \\
             & = & -G \alpha \left(\frac{\dou \psi}{\dou x_3} n_2 - \frac{\dou
\psi}{\dou x_2} n_3 \right) \non \\
             & = & - G \alpha \left(\frac{\dou \psi}{\dou x_3} s_3 + \frac{\dou
\psi}{\dou x_2} s_2 \right) \non \\
             & = & - G\alpha\frac{\dou \psi}{\dou s} \mylabel{fone}            
\eea
The surface stress $\tau_{1s}$ is related to the surface strain as
\bea
\tau_{1s} = 2 S \epsilon_{1s}
\eea
where $S$ is an appropriate surface shear modulus ($\tau^0_{1s}$ can
be taken to be zero with out loss of generality), i.~e.,
\bea
\tau_{1s} = 2 S \epsilon_{1s}& = & 2 S ( -\epsilon_{12} n_3 + \epsilon_{13} n_2) \non \\
          & = & 2 S ( -\frac{\sigma_{12}}{2 G} n_3 +
                       \frac{\sigma_{13}}{2 G} n_2 ) \non \\
          & = &  -S \alpha \left(\frac{\dou \psi}{\dou x_2} n_2 +
\frac{\dou \psi}{\dou x_3} n_3 \right) \non \\
          & = & - S \alpha \frac{\dou \psi}{\dou n} \mylabel{tauones}
\eea
Substituting \prn{fone} and \prn{tauones} in \prn{tauequil} the
boundary condition on $\cal S$ for $\psi$ is obtained as
\bea
\frac{\dou}{\dou s} \left( \psi + \frac{S}{G} \frac{\dou \psi}{\dou n}
\right)= 0, \;\;\;\; \Longrightarrow \;\;\;\;  \psi + \frac{S}{G}
\frac{\dou \psi}{\dou n} = C
\eea
where $C$ is a constant along the boundary. It is assumed that $S$
does not depend on $s$, i.e., all the bounding planes are assumed to
be crystallographically equivalent.

The torsional rigidity of the bar can be computed as follows.
The torque $T$ has contributions from the bulk $\cal B$ and the
surface $\cal S$ and is given as
\bea
T & = & \int_{\cal B} (x_2 \sigma_{13} - x_3 \sigma_{12}) \D{x_2} \D{x_3}
+ \int_{\cal S} \tau_{1s} (x_2 n_2 + x_3 n_3) \D{s} \non \\
  & = & 2 G \alpha \int_{\cal B} \psi \D{x_2} \D{x_3} - G \alpha
\int_{\cal S} \left(\psi + \frac{S}{G} \frac{\dou \psi}{\dou n} \right) \D{s}.
\eea
Evidently, the constant $C$ can be chosen to be zero. The
torsional rigidity  $D$ can be obtained by solving the {\em mixed}
boundary value problem
\bea
\frac{\dou^2 \psi}{\dou x_2^2} + \frac{\dou^2 \psi}{\dou x_3^2}& = & -2
\;\;\;\;\;\;\;\; \mbox{in} \;\;\;\;\;\; {\cal B} \non \\
\mylabel{bvp} \\
\psi + \frac{S}{G} \frac{\dou \psi}{\dou n} & = & 0 \;\;\;\;\;\;\;\;
\mbox{on} \;\;\;\;\;\; {\cal S} \non
\eea
and substituting for $\psi$ in the expression
\bea
D = 2 G \int_{\cal B} \psi \, \D{x_2} \, \D{x_3}.
\eea
Clearly, the theory reduces to the standard theory of torsion when $S$
is set to zero. It is also clear that the key parameter that
determines the atomistic surface effects that affect the torsional
rigidity is the material length-scale $h_0$ defined by the ratio of the
surface shear modulus $S$ and the bulk shear modulus $G$, i.~e., by
\bea
h_0 = \frac{S}{G}. \mylabel{hzero}
\eea
In the next section, a perturbative solution to the boundary
value problem \prn{bvp} will be developed and a general formula for the
size-dependent torsional rigidity will be obtained.

\subsection{Perturbative Solution}
The general perturbative solution will be developed in a
non-dimensional form. To this end, the geometry of the cross section of the bar is
assumed to be characterised by a length scale $a$ (for example, if the
cross section is a square, $a$ can be chosen as one half of the side
of the square). The following non-dimensional quantities are introduced
\bea
\Psi  =   \frac{\psi}{a^2}, \;\;\;\; \xi  =  \frac{x_2}{a}, \;\;\;\; \eta  =  \frac{x_3}{a}. 
\eea
In terms of these nondimensional quantities, the boundary value
problem \prn{bvp} can be recast as
\bea
\nabla^2 \Psi & = & -2
\;\;\;\;\;\;\;\; \mbox{in} \;\;\;\;\;\; {\cal B} \non \\
\mylabel{ndbvp} \\
\Psi + \beta \, \frac{\dou \Psi}{\dou n} & = & 0 \;\;\;\;\;\;\;\;
\mbox{on} \;\;\;\;\;\; {\cal S} \non
\eea
where $\nabla^2 = \frac{\dou^2 }{\dou \xi^2} + \frac{\dou^2 }{\dou \eta^2}$ and the rigidity is
\bea
\frac{D}{G a^4} = 2 \int_{\cal B} \Psi \, \D{\xi} \, \D{\eta}, \mylabel{ndtrgd}
\eea
with 
\bea
\beta = \frac{h_0}{a} \mylabel{beta},
\eea
the nondimensional parameter that governs the extent of surface
effects. When $\beta = 0$, the boundary value problem is solved by the
nondimensional stress function $\Psi_0$ and the torsional rigidity is 
\bea
\frac{D_c}{G a^4} = 2 \int_{\cal B} \Psi_0 \, \D{\xi} \, \D{\eta}
\eea
where the subscript $c$ is used to denote that standard continuum
value of the torsional rigidity. 

When $\beta \ne 0$, the nondimensional stress function can be expanded
in a perturbative series in $\beta$ as
\bea
\Psi = \Psi_0 + \beta \, \Psi_1 + \beta^2 \, \Psi_2 + \ldots \mylabel{perturb}
\eea
Since $\Psi$ solves the boundary value problem \prn{ndbvp}, it follows
that 
\bea
\beta \, \nabla^2 \Psi_1 + \beta^2 \, \nabla^2 \Psi_2 + \ldots = 0 \mylabel{pbvp}
\eea
in $\cal B$ and
\bea
\beta \, \left(\Psi_1 + \frac{\dou \Psi_0}{\dou n} \right) + \beta^2 \,
\left(\Psi_2 + \frac{\dou \Psi_1}{\dou n} \right) + \ldots = 0 \mylabel{pbc}
\eea
on $\cal S$. Since the perturbative expansion \prn{perturb} is valid
for all values of $\beta$, \prn{pbvp} and \prn{pbc} imply that 
\bea
\nabla^2 \Psi_1 & = & 0 \;\; \mbox{on} \;\; {\cal B} \;\; \mbox{with} \;\;
\Psi_1 = - \frac{\dou \Psi_0}{\dou n} \;\; \mbox{on} \;\; {\cal S}
\mylabel{pI} \\
\nabla^2 \Psi_2 & = & 0 \;\; \mbox{on} \;\; {\cal B} \;\; \mbox{with} \;\;
\Psi_2 = - \frac{\dou \Psi_1}{\dou n} \;\; \mbox{on} \;\; {\cal S}
\mylabel{pII} \\
    & . & \non \\
    & . & \non \\
    & . & \non \\
\nabla^2 \Psi_k & = & 0 \;\; \mbox{on} \;\; {\cal B} \;\; \mbox{with} \;\;
\Psi_k = - \frac{\dou \Psi_{k-1}}{\dou n} \;\; \mbox{on} \;\; {\cal S}
\mylabel{pk} 
\eea
which is a sequence of Dirichlet boundary value problems. Thus, the
solution of the boundary value problem \prn{ndbvp} can be obtained to
any desired accuracy in the parameter $\beta$, and the torsional
rigidity obtained as
\bea
\frac{D}{G a^4} = A_0 + A_1 \, \beta + A_2 \, \beta^2 + \ldots
\eea
where 
\bea
A_k = 2 \int_{\cal B} \Psi_{k} \, \D{\xi} \, \D{\eta}
\eea
are constants that depend {\em only on the shape of the
cross-section}. Since $D_c/G a^4 = A_0$, a general expression for the
size dependence of the torsional rigidity can be derived as
\bea
\frac{D - D_c}{D_c} = \frac{A_1}{A_0} \, \beta + \frac{A_2}{A_0} \,
\beta^2 + \ldots \mylabel{sdrigid}
\eea 
For cross-sections of a given shape, the constants $A_k$ can be
calculated once and for all and the formula \prn{sdrigid} along with
\prn{hzero} and \prn{beta} can be used to predict the size dependence
of the torsional rigidity. Atomistic inputs are required only in
providing values for $S$ and $G$. It will be seen in the following
sections that the value of $\beta$ in real systems is less that 0.5
and the most important contribution is from the first term on the right
hand side of \prn{sdrigid}, i.~e.,
\bea
\frac{D - D_c}{D_c} \approx \frac{A_1}{A_0} \, \beta = \frac{A_1}{A_0} \,
\frac{h_0}{a}
\eea

\section{Atomistic Simulations}
\label{Simulations}
Atomistic simulations are carried out to validate the theory developed
in the previous sections.  The bars selected for study
have a square cross section (\fig{torschm}) made of selected FCC metals
(Al, Ag, Cu, Ni) such that the `1'-direction corresponds to the
$[100]$ crystallographic direction and `2' and `3' directions
correspond to $[010]$ and $[001]$ crystallographic directions
respectively, i.~e., the bounding free surfaces are  planes of the
$\{100\}$ family. 

Two sets of simulations are carried out. In the first set, the
constitutive constants $G=C_{44}$ (bulk shear modulus) and $S$ the
surface elastic constant for surface shear of the $\{100\}$ surface are
evaluated atomistically; these are the parameters required as input to
the theory developed. In the second set of simulations the torsional
rigidity is directly calculated using atomistic simulation of
nanoscale torsion. These results are then compared with the
theoretical results of torsional rigidity in the next section.

The atomistic model used in the present study is the embedded atom
method (EAM)  developed by \citeasnoun{Daw1984}. The elements Ag,
Cu, and Ni are modeled using the EAM potentials of
\citeasnoun{Oh1988} and Al is modeled with potentials developed by
\citeasnoun{Ercolessi1994}.  

\subsection{Determination of Surface Elastic Constants}
The surface elastic constants are determined as follows. A block of
atoms are stacked in an FCC crystal lattice such that the coordinate
planes are of the $\{100\}$ family. Periodic boundary conditions are
imposed in the `1' and `2' directions alone simulating bounding free
surfaces of $\{100\}$ family in the `3' direction. The positions of the
atoms are changed to  correspond to a deformation gradient
tensor $\bF = \bI + \epsilon \be_1 \otimes \be_2$ 
that produces a simple shear by amount $\epsilon$ ($\bI$ is the
identity tensor, and $\be_i$ are the orthonormal basis vectors). The
potential energy of this atomistic system is minimised and the
minimised total energy is calculated. To compute the surface energy,
the elastic energy stored in the bulk is subtracted from the
calculated total energy. On performing
this simulation for various values of $\epsilon$, the surface energy
$\gamma$ is obtained as a function of $\epsilon$, and the surface
elastic constant is calculated by numerical differentiation of this
function. It is noted that to compute the bulk elastic constant
$G(=C_{44})$ a similar procedure can be adopted with periodicity
imposed in the `3' direction as well. The results of these simulations
are shown in table 1.

\begin{table}
\label{smodtable}
\bcen
\btab{||c||c||c||c||c||}
\hline
Material & $a_0\,$(\AA) & $C_{44}\,$(eV/\AA$^3$) & $S\,$ (eV/\AA$^2$)
& $h_0\,$ (\AA) \\
\hline
\hline
Al &   4.032 &   0.229 &   0.481 &   2.099 \\
Ag &   4.090 &   0.292 &   0.248 &   0.849 \\
Cu &   3.615 &   0.474 &   0.429 &   0.906 \\
Ni &   3.520 &   0.808 &   0.763 &   0.945 \\
\hline
\etab
\ecen
\caption{Properties of materials calculated using EAM
potentials. $a_0$--lattice parameter, $C_{44}$--bulk shear modulus,
$S$--surface shear modulus and $h_0 = S/C_{44}$.}
\end{table}

\subsection{Atomistic Simulations of Torsional Response}
\begin{figure}
\centerline{\epsfxsize=12.5truecm \epsfbox{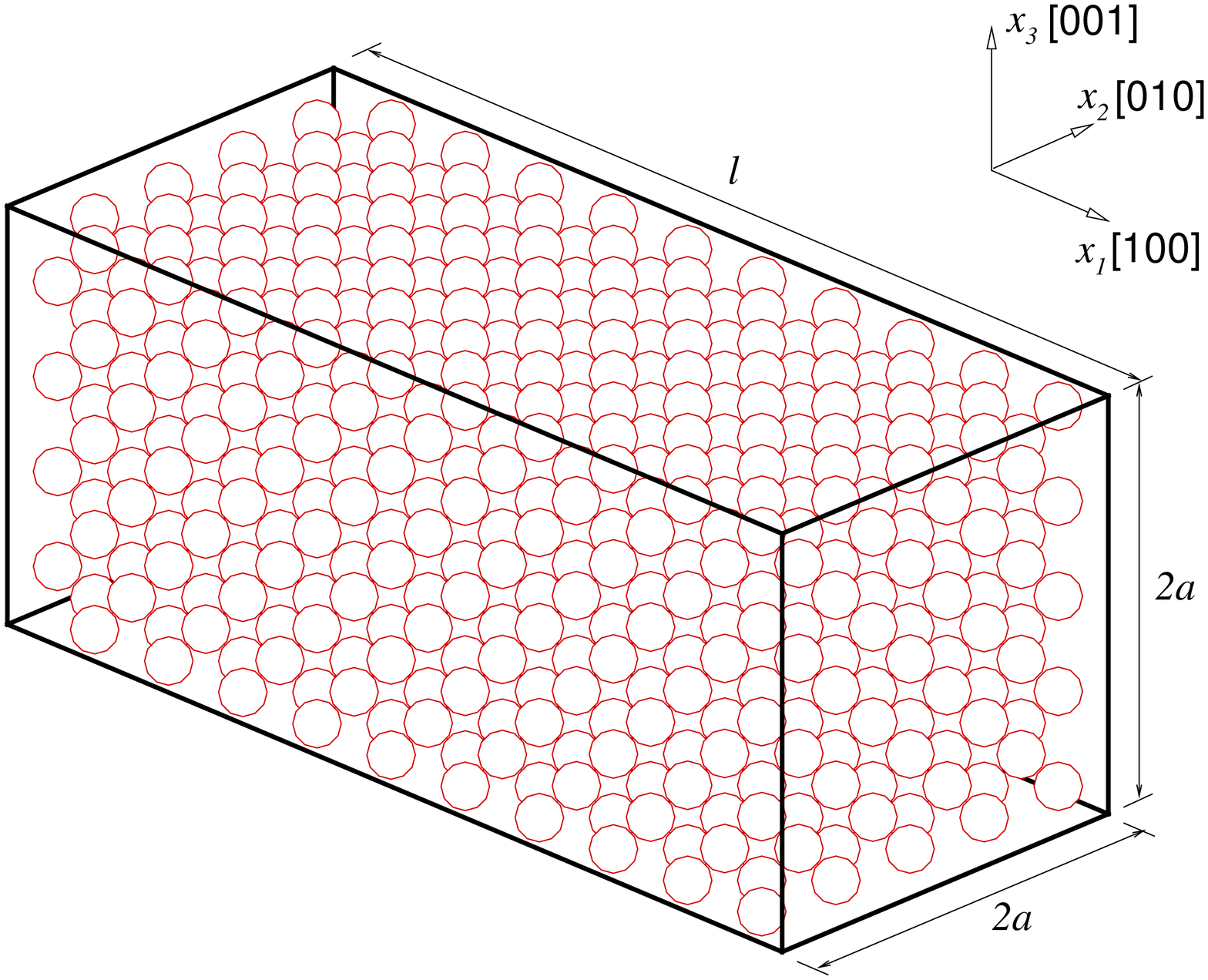}}
\caption{Atomistic system used for the simulation of torsional
response.}
\label{torschm}
\end{figure}

Atomistic simulations used to calculate the torsional rigidities
are performed as follows. The simulation box $(0\le x_1 \le \ell,|x_2|
\le a, |x_3| \le a)$ consists of a collection
of atoms as shown in \fig{torschm} with periodic boundary conditions
imposed in the `l' direction with a periodic length of $\ell$. The
atoms are then displaced according to the rotation tensor 
\bea
\bR(x_1) = \bI + \alpha\, x_1 \, (\be_3 \otimes \be_2 - \be_2
\otimes \be_3)
\eea
where $\alpha$ is the twist per unit length ($\alpha$ is chosen such
that $\alpha \ell \ll 1$, so that the infinitesimal rotation tensor is
sufficiently accurate). With this initial condition, the atomistic
potential energy of the system is minimised. The energy and force
computation during the minimisation process is carried forth as
follows. To compute the energy of an atom (say $i$) near $x_1 =
\ell$ the position vector of its periodic neighbour (say $j$), found
near $x_1 = 0$ in the box, is required. The position vector of the
periodic image of atom $j$ is obtained by applying the rotation tensor
$\bR(\ell)$ to the vector $\bx(j) + \ell \be_1$ where
$\bx(j)$ is the position vector of atom $j$ in the simulation
box. Similarly the periodic neighbour $j$ of an atom $i$ near $x_1 = 0$
can be obtained by applying the inverse of the rotation tensor
$\bR(\ell)$. This procedure ensures that the periodicity in the `1'
direction is maintained, while keeping the bar in a twisted position
with twist per unit length $\alpha$.

Using the procedure above, simulations are carried out for various
values of $\alpha$ and the minimised atomistic total  energy  $E$ in
the simulation box is obtained as a function of $\alpha$. The
torsional rigidity is calculated using the formula
\bea
D = \frac{1}{\ell} \frac{\dou^2 E}{\dou \alpha^2}.
\eea 
The correctness of this procedure is ascertained by choosing various
values of $\ell$ for the periodic distance and computing $D$ using the
above procedure. It is found that $D$ is insensitive to the
choice of $\ell$.

\section{Results and Discussion}
\label{Results}
\subsection{Theoretical Results for Square Bars}
\subsubsection{Exact Solution}
The exact solution of the boundary value problem \prn{bvp} for the
case of a square bar is obtained in this section. The square section
is assumed to be of side $2a$ and the solution for the boundary value
problem is obtained in the nondimensional form \prn{ndbvp} where the
square occupies the region $|\xi| \le 1$, $|\eta| \le 1$. A
straightforward analysis gives that
\bea
\Psi(\xi,\eta) = 8 \sum_{n=1}^{\infty} \frac{\sin{k_n}}{k_n^2 (2
k_n + \sin{2k_n})} \, \cos{k_n \xi} \, \left(1 - \frac{\cosh{k_n
\eta}}{\cosh{k_n} + \beta k_n \sinh{k_n}} \right) \mylabel{exact}
\eea
where $k_n$ is the $n$-th root of the equation
\bea
\cos{k} - \beta k \sin{k} = 0.
\eea
The nondimensional theoretical torsional rigidity \prn{ndtrgd} is
obtained as
\bea
\frac{D}{G a^4} = 64 \sum_{n=1}^{\infty}  \frac{\sin^2{k_n}}{k_n^3 (2
k_n + \sin{2k_n})} \, \left(1 - \frac{\sinh{k_n}}{k_n(\cosh{k_n} +
\beta k_n \sinh{k_n})} \right). \mylabel{exttrgd}
\eea

For the purpose of comparison with atomistic simulations, the
nondimensional warping function $\Phi = \phi/a^2$ is also derived:
\bea
\Phi(\xi,\eta) = \xi\, \eta - 8
\sum_{n=1}^{\infty} \frac{\sin{k_n}}{k_n^2 (2 k_n +
\sin{2k_n})(\cosh{k_n} + \beta k_n \sinh{k_n})} \, \sin{k_n \xi}
\sinh{k_n \eta} \label{thwarp}
\eea

\subsubsection{Perturbative Solution}
The function $\Psi_0$ is obtained by setting $\beta = 0$ in the exact
solution \prn{exact}. The function $\Psi_1$ is obtained by solving the
Dirichlet problem \prn{pI} as
\bea
\Psi_1(\xi,\eta) = 4 \sum_{n=0}^{n=\infty} \frac{\sin{q_n}}{q_n^2}
\frac{\tanh{q_n}}{\cosh{q_n}} \, \left( \cos{q_n \xi} \cosh{q_n \eta}
+  \cosh{q_n \xi} \cos{q_n \eta} \right)
\eea
where $\displaystyle{q_n = \frac{(2\,n -1) \pi}{2}}$. The constant
$A_1$ can be calculated as
\bea
A_1 = 64 \sum_{n=1}^{\infty} \frac{\sin^2{q_n} \tanh^2{q_n}}{q_n^4}
\eea

The numerical values of $A_0$ (which is calculated from \prn{exttrgd}
with $\beta = 0$) and $A_1$ are 
\bea
A_0 = 2.2492\;\;\;A_1 = 8.9969
\eea 
Thus,
\bea
D \approx A_0 + A_1 \beta \label{ptbtrgd}
\eea
and
\bea
\frac{D-D_c}{D_c} \approx \frac{A_1}{A_0} \, \beta = 4\,\beta = 4 \,
\frac{h_0}{a}. \label{dmdc}
\eea

\begin{figure}
\centerline{\epsfxsize=12.0truecm \epsfbox{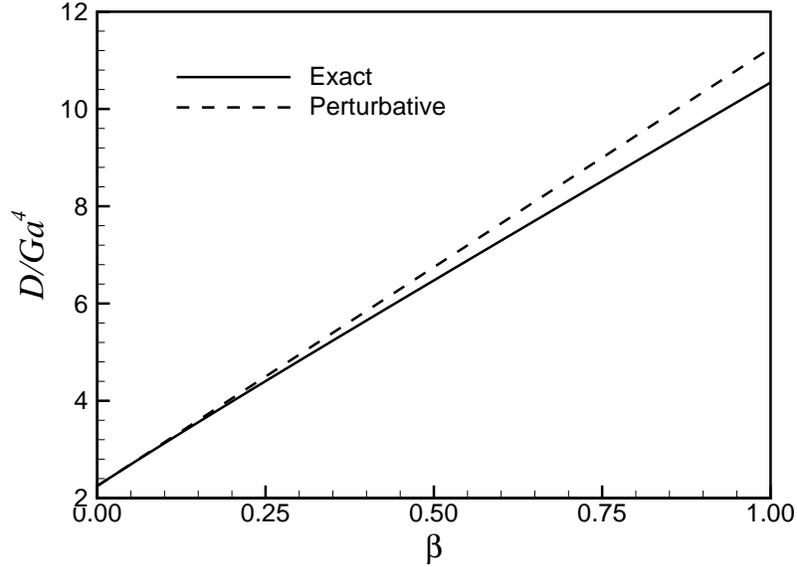}}
\caption{Comparison of the exact \protect{\prn{exttrgd}} and
perturbative \protect{\prn{ptbtrgd}} results for the torsional
rigidity.}
\label{extptb}
\end{figure}

A comparison between the exact result \prn{exttrgd} and the
perturbative result \prn{ptbtrgd} is shown in \fig{extptb}. It is
evident that the perturbative result is a very good approximation of
the exact result; even when $\beta = 1$ (which is much larger than
 that which would appear in systems considered here), the error is
only 6.5\%. Thus the perturbative result in the form of \prn{dmdc} is
used in comparisons with atomistic results.

\subsection{Comparison of Atomistic and Theoretical Results}

\begin{figure}
\centerline{\epsfbox{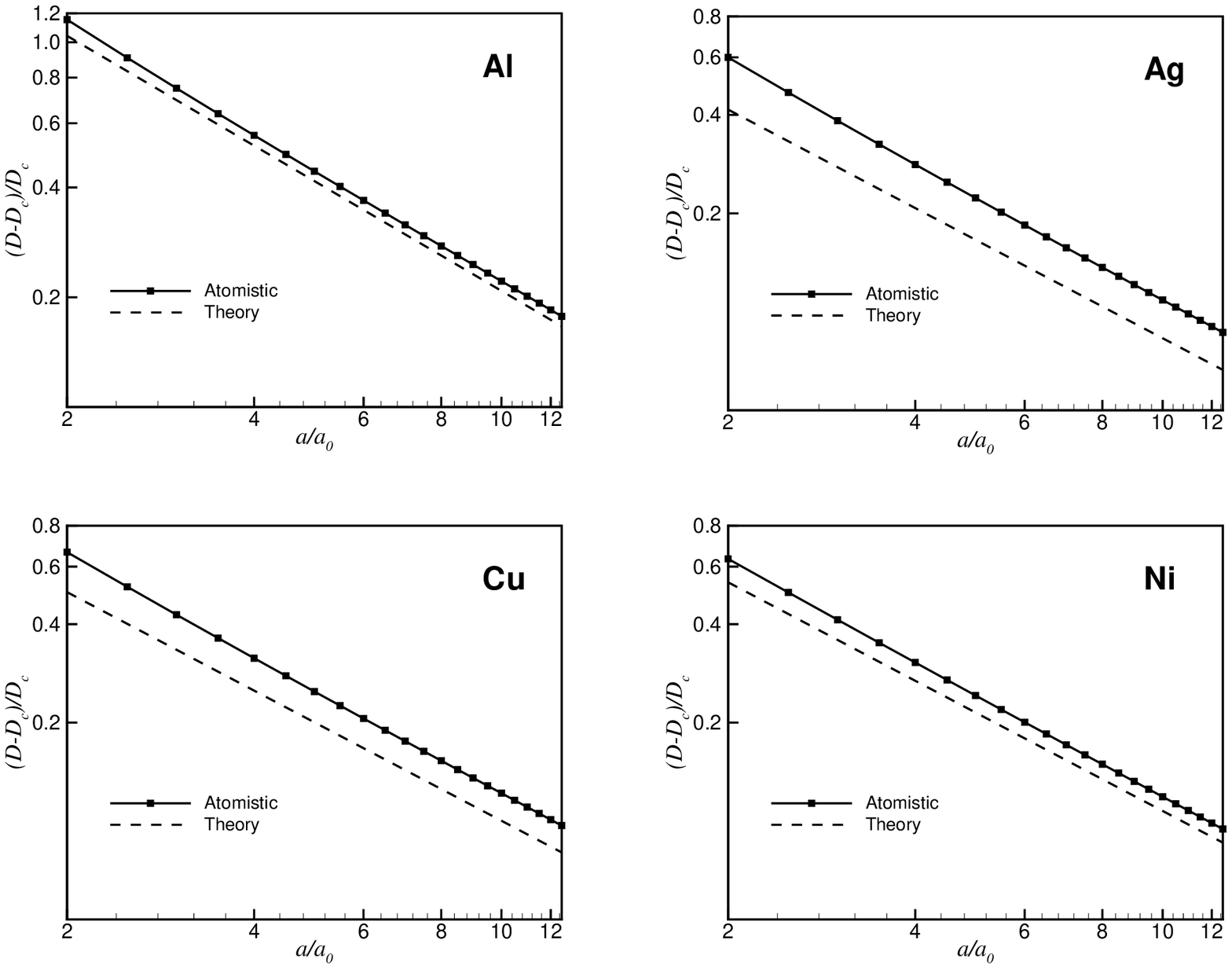}}
\caption{Comparison of the results of atomistic simulations with
theoretical results. $a_0$ is the lattice parameter.}
\label{compare}
\end{figure}

\Fig{compare} shows a comparison between the nondimensional torsional
rigidity computed atomistically and that predicted by the augmented
continuum theory \prn{dmdc}. Several points may be noted:
\begin{enumerate}
\item{It is clear that the atomistically calculated torsional rigidity
differs significantly from the predictions of standard continuum
theory. In fact, for an Al bar of width 5nm the torsional rigidity is
about 50\% larger.}
\item{The nondimensional difference in torsional rigidity computed
atomistically scales very closely as $(1/a)$ in all the metals.}
\item{The atomistically computed values are  accurately predicted
by the theoretical values for Al and Ni (within 10\%), while the
agreement is fair for Ag and Cu (within 30\%).}
\item{In all cases, the atomistically computed values are larger than
the theoretical values, although to varying degrees in different
metals.}
\end{enumerate}

It may be argued that the reason for the atomistic values being greater
than the theoretical values is the neglect of corner effects in the theoretical analysis. A simple dimensional analysis 
indicates that corner effects must scale as $1/a^2$; but the
atomistic results scale very closely as $1/a$. Thus it is clear that
corners play a secondary role in the systems considered here. A more
plausible reason for the difference in the theoretical values and
simulation results is the assumption that the surface energy $\gamma$
depends only on surface strain. In reality the surface energy can also
depend on the {\em surface curvature strain} $b_{\alpha \beta}$,
i.~e., the difference between the deformed curvature and the original curvature
of the surface. Thus the surface energy must be a function of both the
surface strain tensor and the surface curvature strain tensor;
mathematically, $\gamma = \gamma(\epsilon_{\alpha \beta}, b_{\alpha
\beta})$. This idea can be expressed more physically; in the present
model, the surface is treated as a membrane, while a more physically
realistic model will be that of a shell with bending
stiffness. Additional evidence in support of this argument is that in the case of
bending of plates treated by \citeasnoun{Miller2000} the augmented
continuum theory differs by the order of 30\% with the atomistic
simulations. The plates treated did not have corners and this effect
can only be attributed to the neglect of the bending energy of the
surfaces.  Clearly this problem requires a  more elaborate
theoretical framework and will be taken up for study in future.

\begin{figure}
\centerline{\epsfysize=12.0truecm \epsfbox{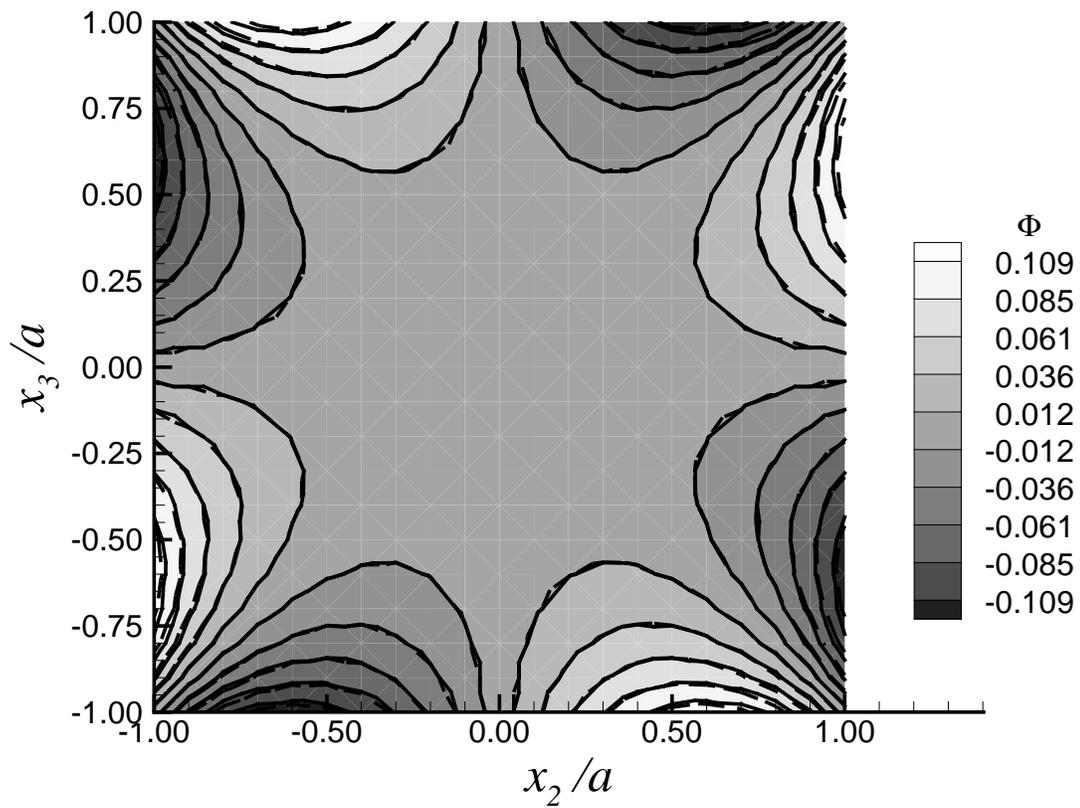}}
\caption{Comparison of atomistically simulated and theoretical
warping function $\Phi$. The solid lines are the contours of the
atomistic result while the dashed lines correspond to the theoretical
calculation \protect{\prn{thwarp}}. The atomistic result is for Al with $a/a_0 = 5$ and the
theoretical result corresponds to $\beta = 0.1$. }
\label{warp}
\end{figure}

The atomistic simulations of torsion not only allow for the
computation of the torsional rigidity but also provide data for the
warping of atomic planes. From the positions of atoms at the
configuration of minimum energy, the values of nondimensional warping
$\Phi$ are calculated. The atomistically computed values of the
nondimensional warping must be independent of the value of $\alpha$
(twist per unit length) according to  the theory. This is indeed found
in the simulations, and provides a further check for the theory.  The
atomistically simulated warping displacements are then compared with
the theoretically predicted values \prn{thwarp}. The result of such an
exercise is plotted in \fig{warp}. It is evident that the predicted
warping is in excellent agreement with the atomistic result.

\section{Conclusions}
\label{Conclusions}

A general framework for the prediction of rigidities of nanoscale
structural elements has been developed and applied to the case of the
nanoscale bars in torsion. The key premise of this theory is that the
heterogeneities present in such systems can be modeled as surface
effects in an augmented continuum theory. A development of this idea
reveals that the material length scale that governs the size
dependence of the rigidity is the ratio of the surface elastic
constant to the bulk elastic constant.  The augmented theory of
torsion developed here is compared with direct atomistic calculations
of bars of various metals and is found to be satisfactory. The author
is not
aware of any work that reports experimentally measured torsional rigidities of
nanosized bars -- indeed, atomistic simulations are used as
numerical experiments. Given the advances in nanotechnology, such
experiments are expected to be performed in the near future.

The use of this theory is envisaged as follows. The bulk
elastic constants and the surface elastic constants (for various
surfaces) of materials of interest can be calculated and
tabulated. The expressions for the constants that appear in the
perturbative expansion in size dependence of the torsional rigidity can
be worked out for a host of cross sectional shapes once and for all.
 A collection of
such information will be useful for the designers of nanomechanical
systems in that the need for direct atomistic simulations of nanoscale
structures is obviated. The work of \citeasnoun{Miller2000} along with the
present work  provide a complete framework for the
prediction of rigidities of nanoscale structural elements in
extension, flexure and torsion. 

Several points for future work are noted. The atomistic model used in
this study, EAM, is known to be inaccurate with applied to interfacial
properties. This, of course, does not invalidate the present work
since the parameters used in the theory and the simulation results are
obtained from the same EAM model. To obtain accurate values of surface
elastic constants more sophisticated atomistic models such as density
functional theory (DFT) may be applied. Another important point to be
noted is that thermal effects are not accounted in the calculation of
the surface elastic constants. Also, the model developed here treats
the surface as a membrane while a more realistic model will have to
additionally account for the bending stiffness of the surface. Results
of investigations along the afore mentioned directions will be
reported in future publications.

\subsection*{Acknowledgement}
The author wishes to thank Ron Miller and Rob Phillips for useful
discussions. Partial support for this work provided by CDAC, Poona,
India, is acknowledged.

\bibliographystyle{ijss}
\bibliography{torrefs.bib} 

\end{document}